   \documentstyle[aps,prl,twocolumn,epsfig]{revtex}

\begin{document}
   \draft

   \twocolumn[\hsize\textwidth\columnwidth\hsize\csname @twocolumnfalse\endcsname%

   \title{Scaling detection in time series: diffusion entropy analysis}
   \author{Nicola Scafetta$^{1}$,    and Paolo Grigolini$^{1,2,3}$.}
   \address{$^{1}$Center for Nonlinear Science, University of North Texas,
   P.O. Box 311427, Denton, Texas 76203-1427 }
   \address{$^{2}$Dipartimento di Fisica dell'Universit\`a di Pisa and
   INFM, Piazza Torricelli 2, 56127 Pisa, Italy}
   \address{$^{3}$Istituto di Biofisica CNR, Area della Ricerca di Pisa,
   Via Alfieri 1, San Cataldo 56010 Ghezzano-Pisa, Italy}
   \date{\today}
   \maketitle

   \begin{abstract}
  The methods currently used to determine the scaling exponent of a complex dynamic process
described by a time series are based on the numerical evaluation of variance. This means that
all of them can be safely applied only to the case where ordinary
statistical properties hold true even if strange kinetics are involved. We illustrate a  method of statistical analysis based on the Shannon entropy
of the diffusion process generated by the time series, called Diffusion Entropy Analysis (DEA).  We adopt artificial Gauss and L\'{e}vy time series, as prototypes of ordinary and anomalus statistics, respectively, and we analyse them with the DEA and four ordinary methods of analysis, some of which are very popular.
We show that the DEA  determines the correct scaling exponent even when the statistical properties, as well as the dynamic properties, are anomalous. The other four methods produce correct results in the Gauss case but fail to detect the correct scaling in the case of L\'{e}vy statistics.

   \end{abstract}
   \vspace{0.5cm}
   %
   ] 
   %
\section{Introduction}

Scale invariance has been found to hold empirically for a number of complex systems
and the correct evaluation of the scaling exponents is of fundamental importance
to assess if universality classes exist \cite{stanley}. The mathematical definition of scaling is as follows. The function $\Phi(r_1,r_2,\ldots)$ is termed  scaling invariant, if it fulfills the property:
\begin{equation}
\label{scafun11}
\Phi (r_1,r_2,\ldots)=\gamma^a ~\Phi (\gamma^b r_1,\gamma^c r_2,\ldots)~.
\end{equation}
Eq. (\ref{scafun11}) means that if we scale all coordinates  $\{r\}$ 
by means of an appropriate choice of the exponents $a,b,c....$, then we
always recover the same scaling function. The theoretical
and experimental search for the correct scaling exponents is intimately
related to the discovery of deviations from ordinary statistical mechanics. This aspect emerges clearly, for instance, from Ref. \cite{goldenfeld}. The author of this interesting book, with the help of dimensional analysis and regularity assumption, determines the values of the scaling exponents. These scaling exponents, however, are \emph{trivial} in the sense that they refer to ordinary statistical mechanics.

In this paper we focus on the scaling of time series, and consequently \cite{dfa} on the scaling of diffusion processes. In fact, according to the prescription of Ref. \cite{dfa} we interpret the numbers of a time series as generating diffusion fluctuations, thereby shifting our attention  from the time series to the probability distribution function (pdf) $p(x,t)$, where $x$ denotes the  variable collecting the fluctuations. In this case, if the time series is stationary, the scaling property of Eq. (\ref{scafun11}) takes the form
\begin{equation}\label{scafun12}
p(x,t) = \frac{1}{t^{\delta}}~F\left( \frac{x}{t^{\delta}}\right),
\end{equation}
where $\delta$ is the scaling exponent. The ordinary value  \cite{goldenfeld}  of the scaling exponent is $\delta = 1/2$. Ordinary statistical mechanics is intimately related to the Central Limit Theorem (CLT) \cite{khinchin}, thereby implying for for the function $F$ of Eq. (\ref{scafun12}) the Gaussian form.

The main purpose of this paper is to prove that a technique of statistical analysis, recently introduced to establish the thermodynamic nature of a time series of sociological interest \cite{nicola}, affords a reliable way to  evaluate the scaling exponent. This method of analysis is based on the entropy of the diffusion process and for this reason is called Diffusion Entropy Analysis (DEA). We compare the DEA  to  the Standard Deviation Analysis (SDA) \cite{barbi},  to the Detrended Fluctuation Analysis (DFA) \cite{dfa},  to the Rescaled Range Analysis (RRA) \cite{hurstbook},  to the  Spectral Wavelet Analysis (SWA) \cite{percival}, and we show that, while all these techniques, some of which are very popular, can yield wrong scaling exponents, the DEA always determines the correct value, with satisfactory precision. This important conclusion is reached by examining artificial sequences generating both Gauss and L\'{e}vy statistics. The DEA is the only technique yielding the correct scaling in both cases. The other techniques produce correct results only  in the Gauss case but fail to detect the correct scaling in the case of L\'{e}vy statistics.

  \section{DEA}
It is remarkably simple to determine the scaling parameter $\delta$ using  DEA.  First of all, we transform the time series into a diffusion process with a given pdf $p(x,t)$ (Section 4 illustrates an algorithm to do that).  Then, we measure the Shannon entropy
\begin{equation}\label{scafun13}
S(t) = - \int_{-\infty}^{+\infty} dx~ p(x,t)~ \ln \left[p(x,t)\right].
\end{equation}
Let us suppose that $p(x,t)$ fits the scaling condition of 
Eq.(\ref{scafun12}) and let us plug Eq.(\ref{scafun12}) into 
Eq.(\ref{scafun13}).
After an easy algebra, based on changing the integration variable from $x$ into $y = x/t^{\delta}$,
we obtain 
\begin{equation}\label{scafun14}
S(\tau) = A + \delta~ \tau,
\end{equation}
	    where
	    \begin{equation}
		A \equiv -\int_{-\infty}^{\infty} dy \, F(y) \, \ln [F(y)],		\label{ainthecontinuouscase}
		\end{equation}
and
\begin{equation}\label{logarithmictime}
\tau  \equiv \ln (t).
\end{equation}

It is evident that the origin of this logarithmic time, $\tau = 0$,  is related to $t = 1$, and this, in turn, depends on the time unit adopted.  However, this arbitrary choice in no way affects the scaling parameter, which
is given by the slope of the straight line $S(\tau)$ of Eq. (\ref{scafun13}).  The adoption of a different
time unit would change a given straigh line into a new one,  parallel  to the original, and thus bearing the same scaling parameter $\delta$.

\section{Gauss and L\'{e}vy Diffusion}

The rigorous definition of diffusion scaling is that of Eq.(\ref{scafun12}). Frequently, the scaling property is also expressed
by means of
\begin{equation}
    x \propto t^{\delta}.
 \label{scalingdefinition2}
 \end{equation}
Let us see now why  the scaling paramater $\delta$ is often evaluated
through the second moment of the pdf $p(x,t)$. In the long-time limit the variable $x(t)$ collecting the fluctuations
$\xi(t)$ has a time evolution equivalent to

\begin{equation}\label{dynammodeflw}
\stackrel{{\bf \cdot} }{x}(t)=\xi (t)~,
\end{equation}
By time integration we get
\begin{equation}
\label{soldiffdynalw}
x(t)=x(0)+\int\limits_{0}^{t}dt'~\xi (t')~.
\end{equation}
Let us imagine a set of infinitely many trajectories of the type of that of Eq. (\ref{soldiffdynalw}). As to the second moment $<x^{2}(t)>$, we evaluate its time evolution by squaring Eq.(\ref{soldiffdynalw}) and averaging over all the trajectories of this set. Under the assumption that the process
is stationary and that $<\xi(t)> = 0$, it is straightforward to obtain
\begin{equation}
<x^{2}(t)> = <x^{2}(0)> + 2 <\xi^{2}> \int_{0}^{t} dt_{1} \int_{0}^{t_{1}}  dt_{2} \Phi_{\xi}(|t_{2}| ).
\label{secondmoment}
\end{equation}
Note that to get this result we use the equilibrium correlation function 
\begin{equation}\label{norcorrfun23f}
\Phi_{\xi} (t_1,t_2) \equiv \frac{<\xi(t_1)\xi(t_2)>}{<\xi^2>}.
\end{equation}
Under the stationary condition, this correlation function depends only on the time difference, namely,
$\Phi_{\xi} (t_1,t_2) =\Phi_{\xi} (|t_1-t_2|) $, and this property, with the help of a suitable change of integration variables,
yields Eq.(\ref{secondmoment}). 

What is the connection between second moment and scaling? Having in mind Eq.(\ref{scalingdefinition2}) one would be tempted to make the
conjecture that
\begin{equation}
<x^{2}(t)> \propto t^{2\delta}.
\label{secondmomentscalingd}
\end{equation}
However, this conjecture is not correct in general, and to be more rigorous let us replace
Eq.(\ref{secondmomentscalingd}) with
\begin{equation}
<x^{2}(t)> \propto t^{2 H}.
\label{secondmomentscalingH}
\end{equation}
The adoption of the symbol $H$ rather than the symbol $\delta$ is dictated by two main reasons. First of
all, $H$ is the symbol frequently used in literature to denote the scaling parameter. In fact,  the pioneer work of Mandelbrot, in the special case 
of Fractional Brownian Motion  (FBM) \cite{2Mandelbrot}, 
identified the Hurst coefficient $H$  \cite{hurstbook} with the scaling parameter. This is certainly correct, but the validity of this conclusion is confined to the Gaussian case. In general, as we shall see, the asymptotic behavior of Eq.(\ref{secondmomentscalingH}) does not imply that
$\delta = H$. This is the second reason why we prefer to use Eq.(\ref{secondmomentscalingH}) rather than Eq.(\ref{secondmomentscalingd}). In conclusion, we denote by $\delta$ a parameter implying the property of Eq.(\ref{scafun12}) and by $H$ the value corresponding to the property of Eq.(\ref{secondmomentscalingH}). We shall see that even when $H \neq \delta$, the parameter $H$ might have a physical meaning of some interest,
even if it is not the real scaling.

To prove under which conditions the equality $\delta = H$ applies, and consequently Eq. (\ref{secondmomentscalingd}), as well as
Eq.(\ref{secondmomentscalingH}), is correct, let us notice that under the assumption that the fluctuation $\xi(t)$ is Gaussian, and with no other assumption, we can prove that the pdf $p(x,t)$ fulfills the following diffusion equation 
\begin{equation}
\label{gaussiandiffusioequation}
\frac{\partial p(x,t)}{\partial t} =  D(t) \frac{\partial^{2}}{\partial x^{2}} p(x,t),
\end{equation}
where
\begin{equation}
D(t) \equiv <\xi^{2}> \int_{0}^{t} \Phi_{\xi}(t') dt' .
\label{timedependentdiffusion}
\end{equation}
The proof of this important result rests on the cumulant theory of Ref.\cite{kubo} and the interested reader can derive it from a more general case
discussed in Ref.\cite{mario}. It is straightforward to show that the general solution of Eq. (\ref{gaussiandiffusioequation}), for a set of particles initially located 
at $x = 0$ , is
\begin{equation}
\label{towardsscaling}
p(x,t) = \frac{1}{\sqrt{2 \pi <x^{2}(t)>}} ~\exp\left( -\frac{x^{2}}{2<x^{2}(t)>}\right),
\end{equation}
where $<x^{2}(t)>$ is the second moment with the time evolution described by Eq.(\ref{secondmoment}). It is easy to show that the time asymptotic properties of the second moment are compatible with Eq.(\ref{secondmomentscalingH}), with $H$ ranging from $0$ to $1$, 
in the case where
\begin{equation}
\label{towardsscaling2}
lim_{t \rightarrow \infty} \Phi_{\xi}(t) =  sign \frac{const}{t^{\beta}},
\label{beta}
\end{equation}
with 
\begin{equation}
\label{beta1}
H = 1 - \frac{\beta}{2}
\end{equation}
and $0 \leq \beta \leq1$, if $sign = 1$,
and $1 \leq \beta \leq 2$, if $sign = -1$. 
It is evident that in this physical condition the property of Eq.(\ref{scafun12}) applies to the time asymptotic limit, and consequently
in the same time limit the property $\delta = H$ holds true.
In Section 3A we shall shortly review the FBM, which is a matematical idealization fitting the condition $\delta =H$. In Section 3B we shall illustrate the L\'{e}vy flight diffusion, which is, on the contrary, a mathematical idealization incompatible with the existence of a finite second moment, thereby implying a striking violation of $\delta = H$. Finally, in Section 3C we illustrate the L\'{e}vy walk diffusion, a compromise between the two earlier conditions, in the sense that it is compatible with a finite second moment. 
However, as we shall see, also the L\'{e}vy walk diffusion results in a violation of  $\delta = H$.

\subsection{Fractional Brownian Motion}

Fractional Brownian Motion, FBM, is a generalization of ordinary Brownian motion introduced by Mandelbrot \cite{2Mandelbrot} to describe anomalous
diffusion. By definition, the FBM of index $\eta$ is described by the fractional Gaussian propagator
\begin{equation}\label{gaussdffgfr}
p(x,t)=\frac{1}{\sqrt{4\pi Dt^{\eta}}}~\exp \left( -\frac{(x-\overline{x})^2}{4Dt^{\eta}}\right)~.
\end{equation}
The variance of this pdf is given by
\begin{equation}\label{secoom2fbn}
<(x-\overline{x})^2>=\int\limits_{-\infty }^{\infty }~dx~(x-\overline{x})^2~p(x,t)=2D~t^{\eta}~.
\end{equation}

We see that FBM is compatible with the asymptotic limit of the dynamic approach to diffusion generated by 
Eq. (\ref{dynammodeflw}) when the fluctuation $\xi(t)$ is Gaussian. 
 Furthermore, we see that Eq.(\ref{gaussdffgfr}) fits the scaling condition of Eq. (\ref{scafun12}) and Eq.(\ref{secoom2fbn}) fits the second moment scaling of Eq. (\ref{secondmomentscalingH}). 
Thus we conclude that
\begin{equation}\label{fbmhd}
\delta=\frac{\eta}{2}=H~.
\end{equation}
In other words, FBM is a kind of diffusion where the second moment scaling mirrors correctly the scaling property of Eq.(\ref{scafun12}).

\subsection{L\'{e}vy Flight Diffusion}

L\'{e}vy Flight Diffusion is described, in the symmetric case,  by 
 the characteristic function:
\begin{equation}\label{carlevyflu}
\hat{p}(k,t)=\exp \left(-K^{\alpha } ~t~ |k|^{\alpha }\right)~.
\end{equation}
This type of diffusion is generated by a walker that makes  jumps with lengths determined by
a probability density function, $\lambda(\xi)$, whose  asymptotic behavior function has the following inverse power law form:
\begin{equation}\label{levyasbeaf}
\lambda (\xi)\sim A_{\alpha} ~\sigma^{\alpha} |\xi|^{-1-\alpha}=A_{\alpha} ~\sigma^{1-\mu} |\xi|^{-\mu}.
\end{equation}
This means $|\xi|\gg\sigma$ and $\mu=1+\alpha$.  

 The Fourier inversion of
(\ref{carlevyflu}) can be obtained analytically by making use of
the Fox function \cite{Grigoli23,Jesperrs23}
\begin{eqnarray*}\label{foxlevyfly}
 p(x,t)    =           & &        \frac{1}{\mu-1}~\frac{1}{t^{1/(\mu-1)}}~\left(\frac{|x|}{t^{1/(\mu-1)}} \right)^{-1} \cdot  \\
        & &              H_{2,2}^{1,1}\left[ \frac{|x|}{(K^{\alpha}~t)^{1/(\mu-1)}}~|~{(1,1/{\alpha}),(1,1/2) \atop (1,1),(1,1/2)}\right]~.   
\end{eqnarray*}
It is evident that this expression fits the scaling definition of Eq. (\ref{scafun12}), and that consequently, for $1<\mu<3$,
the scaling coefficient $\delta$ is
\begin{equation}\label{deltalfmkk}
\delta=\frac{1}{\mu-1}~.
\end{equation}
It is important to remark that while the CLT yields Eq.(\ref{scafun12}) with $\delta = \frac{1}{2}$ with $F$ being a Gaussian function, the
Generalized Central Limit Theorem (GCLT)
\cite{gnedenko} yields Eq.(\ref{scafun12}) with $\delta \neq \frac{1}{2}$ and $F$ departing from the Gaussian form. 
This departure from Gaussian statistics has the striking consequence of making the second moment diverge.
In fact, in the asymptotic limit of large $x$'s, we get for the distribution  $p(x,t)$ the following inverse power law form
\begin{equation}\label{asimptotj}
\lim _{|x|\to \infty  }p(x,t) \approx  \frac{1}{t^{1/(\mu-1)}}~\left(\frac{|x|}{t^{1/(\mu-1)}} \right) ^{-\mu}~~~~~\mu<3~.
\end{equation}
This makes the second momernt $<x^{2}>$ diverge. In the more general asymmetric case \cite{gnedenko} similar inverse power law
properties apply to one of the two tails, thereby making the variance  $<(x-\overline{x})^2>$
diverge.  Consequently, the variance method for scaling detection is quite inappropriate in this case.
In the case of real data, the experimental observation is done with a finite number of data. This means that the variance is finite, thereby giving the impression that 
the variance method can be safely adopted as a scaling detector. This would lead to misleading results,
determined by the lack of  accurate statistics rather than by the genuine statistical properties of the system under study.

 \subsection{A compromise: L\'{e}vy Walk Diffusion}
     
Let us now consider another random walk prescription, L\'{e}vy Walk Diffusion \cite{giacomo}. This has to do 
with drawing the random numbers $\tau_{i}$'s, with the probability 
density $\psi(\tau)$ given by
 \begin{equation}
\label{densitydistributionoftau}
\psi(\tau) = (\mu-1) \frac{T^{\mu-1}}{(T  + \tau)^{\mu}}.
\end{equation}
This means that we can build up an infinite sequence of numbers, which 
are then used to make a random walker walk with the following rules. 
At time $t=0$, when the first random number, $\tau_{1}$, is selected, 
we also toss a coin to decide whether the random walker has to move 
in the positive or in the negative direction. The random walker walks 
with a velocity of constant intensity $W$. Thus, tossing a coin 
serves the purpose of establishing whether the velocity of the random 
walker is $W$ (head) or $-W$ (tail).  This condition of uniform motion 
lasts for a time interval of duration $\tau_{1}$.  At the end of this 
condition of uniform motion, a new number, $\tau_{2}$, is randomly 
drawn, and a new velocity direction is established by another coin 
tossing. The time series $\{\tau_{i}\}$ is converted into a time series
$\{\xi_{i}\}$ of either $1$'s  or $-1$'s, as follows. We associate each $\tau_{i}$ 
to a patch of $n_{i}$ identical symbols, the $i$-th patch. The values of these symbols are fixed to be either  $1$'s or $-1$'s, according to the coin
tossing rule. Then we sew the $i$-th patch to the $i-1$-th, on the left, and to the $i+1$-th, on the right, thereby creating a virtually infinite sequence of $\xi_{i}$, with values given by either $1$ or $-1$. It is evident that the variable $x$ collecting the fluctuations $\xi_{i}$ 
corresponds to the earlier walking prescription with $W = 1$.
The error resulting from the
adoption of only the integer part of $\tau_{i}$ is made irrelevant by the fact that the inverse power law nature of Eq. ({\ref{densitydistributionoftau}) with $\mu < 3$ makes statistically significant the numbers $\tau_{i} >> 1$. It is worthe remarking that we shall focus our attention on the condition
$2 < \mu < 3$ .

It is important to stress that the physical condition $\mu > 2$  corresponds to the non-vanishing mean time 
$<\tau>$, whose explicit expression is:
\begin{equation}
    <\tau> = \frac{T}{\mu - 2}.
    \label{meanwaitingtime}
    \end{equation}
    We denote as \emph{event} the joint process of random drawing of a 
    number and of coin tossing. We consider a time scale characterized 
    by the property
    \begin{equation}
	t \gg  \tau .
	\label{coarsegrainingtime}
	\end{equation}
	It is evident that the number of events that occurred prior
	to a given time  $t$ is given by
	\begin{equation}
	n = \frac{t}{<\tau>}.
	\label{numberofevents}
	\end{equation}
Consequently, at a given time $t \gg  <\tau>$, the position $x$ occupied by the random walker
is equivalent to the superposition of many highly correlated fluctuations $\xi_{i}$ or of  $n$ uncorrelated variables $\eta_{i}$, whose modulus is
equal to $W \tau_{i}$, with signs randomly assigned by the coin tossing. Thus, the probability distribution function, $\lambda(\eta)$,
is given by
\begin{equation}
\label{jumpdistribution}
\lambda(\eta) = \frac{1}{2 W}~ \psi\left(\frac{\eta}{W}\right),
\end{equation}
the analytical form of the function $\psi$ being given by Eq.(\ref{densitydistributionoftau}). The variables $\eta_{i}$ can be identified
with the variables $\xi$ of Section 3B.
As a consequence, the asymptotic properties of this probability distribution function are the same as those
of Eq.(\ref{levyasbeaf}). By applying again the GCLT \cite{gnedenko} we obtain the same diffusion process as
that of Eq.(\ref{carlevyflu}), and, of course, the same scaling prescription as that of Eq.(\ref{deltalfmkk}).

This walk prescription was termed Symmetric Velocity Model in Ref. \cite{giacomo} and is a form  of L\'{e}vy walk. The L\'{e}vy walk has been originally introduced \cite{levywalk1,levywalk2,levywalk3}
as a dynamic approach to L\'{e}vy statistics more realistic than the L\'{e}vy flight of Section 3B. The reason for this conviction
is that with the L\'{e}vy walk the walker makes a step of a given length in a time proportional to this length, while
with the L\'{e}vy flight the walker makes instantaneous jumps of arbitrarily length, a property judged, in fact, to be somewhat unrealistic.   In this paper, the L\'{e}vy walk
 serves the very useful purpose of explaining
why the emergence of L\'{e}vy statistics does not imply a total failure of the methods relating scaling to variance. In this
case, in fact, the second moment is finite, and this property does not depend on the lack of sufficient statistics. It depends
on the fact that no jump can occur with a length of intensity larger than $Wt$. In this specific case the renewal theory \cite{geisel}
prescribes that the correlation function $\Phi_{\xi}(t)$ is related to the waiting time distribution $\psi(\tau)$ by the important equation
\begin{equation}
\Phi_{\xi}(t) = \frac{1}{<\tau>} \int_{0}^{+\infty} (t'-t) \psi(t') dt'.
\label{geisel}
\end{equation}
From this important relation, using Eq.(\ref{densitydistributionoftau}), we derive the following analytical expression for $\Phi_{\xi}(t)$:
\begin{equation}
\Phi_{\xi}(t) = \left( \frac{T}{t + T}\right)^{\beta}. 
\label{explicitexpression}
\end{equation}
with
\begin{equation}
\label{meaningofbeta}
\beta = \mu -2.
\end{equation}
At this stage we are equipped to derive the asymptotic properties of the pdf second moment. The existence of the correlation function of
Eq.(\ref{explicitexpression}) allows us to use again Eq. (\ref{secondmoment}) so as to reach quickly the conclusion that
\begin{equation}
H = \frac{4 - \mu}{2}.
\label{wrongscaling}
\end{equation}
This result is, in fact, obtained by plugging Eq.(\ref{meaningofbeta}) into Eq.(\ref{beta1}). There is no reason to identify $H$ with $\delta$, in this case. Rather, if we trust the GCLT and, consequently, the scaling prescription
of Eq.(\ref{deltalfmkk}), we see that $\delta$ is related to $H$ by
\begin{equation}\label{relHdelta34}
\delta =\frac{1}{3-2H}~.
\end{equation}
We shall prove that the DE method detects this correct scaling, whereas, of course, the methods resting on variance cannot, even if
the exponent $H$ they detect has an interesting physical meaning. 

It is worth stressing that the physical meaning of $H$, determined by the variance method, changes from case to case and depends on both
the statistics of the process and the walking rule adopted to change the time series into a diffusion process.
To illustrate this issue, let us multiply  the numbers $\tau_{i}$ by either $1$ or $-1$ by tossing a coin.
Then, let us denote the resulting numbers  by the symbol $\xi$. We obtain a sequence indistinguishable from that discussed
in Section 3B. In this case, as we shall see, in Section 6B, the resulting value of $H = 0.5$ reflects the fact that the numbers $\tau_{i}$ are uncorrelated and, by no means, that the resulting statistics is Gaussian.

\section{The Diffusion Algorithm}

Let us consider a sequence of $M$ numbers
\begin{equation}
    \xi_{i} ,    \quad i = 1,  \ldots , M.
    \label{thesequenceundestudy}
    \end{equation}
    The purpose of the DEA algorithm is to establish the possible 
    existence of a scaling, either normal or anomalous, in the most 
    efficient way as possible without altering the data with any form 
    of detrending. First of all, let us select  an integer number
    $l$, fitting the condition $1 \leq l \leq M .$ 
This integer number will be referred to by us as ``time''. For any given 
	time $l$ we can find $M - l +1$ sub-sequences defined by
	\begin{equation}
	    \xi_{i}^{(s)} \equiv \xi_{i + s}, \quad with  \quad s = 0,  \ldots ,  M-l.
	    \label{multiplicationofsequence}
	    \end{equation}
	    For any of these sub-sequences we build up a diffusion trajectory, $s$, defined by the position 
	    
	    \begin{equation}
	x^{(s)}(l) = \sum_{i = 1}^{l} \xi_{i}^{(s)} 
	= \sum_{i = 1}^{l} \xi_{i+s}. 	
	    \label{positions}
	    \end{equation}

	    Let us imagine this position as that of a sort of Brownian particle that 
	   at regular intervals of time has been jumping forward of 
	   backward according to the prescription of the corresponding 
	   sub-sequence of Eq.(\ref{multiplicationofsequence}). This means 
	   that the particle before reaching the position that it holds at 
	   time $l$ has been making $l$ jumps. The jump made at the
	   $i$-th step has the intensity $|\xi_{i}^{(s)}|$ and is forward or 
	   backward according to whether the number $\xi_{i}^{(s)}$ is 
	   positive or negative. We adopt the paradigm of Brownian motion
for the tutorial purpose of illustrating how the diffusion algorithm work.
Actually, the ultimate task of  this algorithm is to
express in a quantitative way the departure of the observed process
from the statistical properties of ordinary Brownian motion.

	   We are now ready to evaluate the entropy of this diffusion process. 
	   To do that we have to partition the $x$-axis into cells of size 
	   $\epsilon(l)$. When this partition is made, we have to label the 
	   cells. We count how many particles are found in the same cell at a 
	   given time $l$. We denote this number by $N_{i}(l)$. Then 
	   we use this number to determine the probability that a particle 
	   can be found in the $i$-th cell at time $l$, $p_{i}(l)$, by means 
	   of
	   \begin{equation}
	    p_{i}(l) \equiv  \frac{N_{i}(l)}{(M-l+1)} .
	    \label{probability}
	    \end{equation}
	    At this stage the entropy of the diffusion process at the time $l$
	    is determined and reads
\begin{equation}
 S_{d}(l) = - \sum_{i} p_{i}(l) ~\ln [p_{i}(l)].
\label{entropy}
		\end{equation}
		The easiest way to proceed with the 
choice of the cell size, $\epsilon(l)$, is to assume it to be a fraction
of the square root of 
the variance of the 
fluctuation $\xi(i)$, and consequently independent of $l$.

\section{The methods of analysis based on variance}
In this section we review four  methods of time series analysis. The last three methods are very popular,  and are all related,  to  
a somewhat different extent,  to the first, based on the direct  evaluation of variance. 

\subsection{SDA}
The direct evaluation of variance, by means of SDA,  is probably the most natural method of variance detection. This method was used, for instance, in Ref. \cite{barbi}. 
The starting point is given by the diffusion algorithm of Section 4, Eq.(\ref{positions}). All
trajectories start from the origin $x(l=0)=0$.  With increasing time $l$, the sub-sequences generate a diffusion
process. At each time $l$, it is possible to calculate  the
Standard Deviation (SD) of the position of the $N-l$ trajectory
with the 
well known expression:
\begin{equation}\label{varvar33}
SD(l) =\sqrt{\frac{ \sum _{s=1}^{N-l}\left(x^{(s)}(l)-\overline{x}(l) \right)^2}{N-l-1}},
\end{equation}
where $\overline{x}(l)$ is the average of the positions of the
$N-l$ sub-trajectories at time $l$. We note that the prescription
illustrated in Section 4 to define the trajectories of this diffusion process
is based on overlapping windows. This means that the trajectories are not
totally independent of one another. In principle, we
can also adopt a non-overlapping window method.  In this case the
largest trajectory that we can build up with the whole sequence is divided in $M=int(N/l)$
smaller trajectories of size $l$ (with 
$int(x)$ we denote the integer part of $x$). Thus, the
quantity $SD(l)$ can be referred to many trajectories totally independent the one from the other. 
The advantage of using many independent trajectories is balanced, though, by the disadvantage of  statistics  poorer than those
obtained by using  the overlapping window method. Therefore, in this paper we use the method of overlapping windows.

According to the traditional wisdom of the methods based on variance, the existence of scaling is assessed by observing, with numerical methods, the following properties
\begin{equation}
    SD(l) \propto l^{H}.
 \label{scalingdefinition3}
 \end{equation}
The exponent $H$ is interpreted as the scaling exponent. As discussed in Section 3, there is no guarantee that this exponent
coincides with the genuine scaling $\delta$. This is the reason why with all the methods of analysis of this section
we shall use the symbol $H$ to denote the result of the statistical analysis. To assess whether this is the true scaling or not, it is necessary 
to use also the DEA.

\subsection{RRA}

The  RRA was introduced by Hurst in 1965, in the work, {\it Long-Term Storage: An Experimental
Study} \cite{hurstbook},  for the main purpose of studying  the water storage of the Nile river. The
problem was to design a reservoir, which never overflows or
empties, based upon the given record of observed discharges from a
lake. Let us suppose that $\xi_i$ is the amount of water flowing
from a lake to a reservoir for each year. The problem is to
determine the needed capacity of the reservoir under the condition
that each year the reservoir releases a volume of water equal to
the mean influx.  In $\tau$ years, the average influx is
\begin{equation}\label{infaveyea3}
<\xi>_{\tau}=\frac{1}{\tau}\sum _{i=1}^{\tau}\xi_i~.
\end{equation}
The amount of water accumulated in the reservoir in $t$ years is
\begin{equation}\label{requatwat3}
x(t,\tau)=\sum _{i=1}^{t}( \xi_i-<\xi>_{\tau})~.
\end{equation}
The reservoir neither overflows nor empties during the period of
$\tau$ years if its  storage capacity  is larger than the
difference, $R(\tau)$, between the maximum and minimum amounts of
water contained in the reservoir. $R(\tau)$ is
\begin{equation}\label{rRRhurst3}
R(\tau)=\max_{1\leq t\leq \tau} x(t,\tau)-\min_{1\leq t\leq \tau} x(t,\tau)~.
\end{equation}
For getting a dimensionless value, Hurst divided $R(\tau)$ by the
standard deviation $S(\tau)$ of the data during the $\tau$ years:
\begin{equation}\label{standevhur3}
S(\tau)=\sqrt{\frac{1}{\tau} \sum _{i=1}^{\tau} (\xi_i-<\xi>_{\tau})^2} ~.
\end{equation}
Hurst observed that many phenomena are very well described  by the following scaling relation:
\begin{equation}\label{hursrel3}
\frac{R(\tau)}{S(\tau)}\propto \tau^H~.
\end{equation}
The exponent $H$ (denoted by the letter $K$ by Hurst) was called Hurst
exponent, and consequently denoted by the letter $H$, by  Mandelbrot \cite{2Mandelbrot}.  The work of Mandelbrot made the RRA become popular
as a technique of scaling detection. 
In the case of the Nile river, Hurst measured an exponent $H=0.9$.
 This means that the Nile is characterized by a long range persistence that requires
 unusually high barriers, such as the Asw\^{a}n High Dam, to contain damage and rein in the floods.
This paper proves that this perspective, correct in the Gaussian case, is in general misleading.

\subsection{DFA}

The DFA was introduced in 1994 by the authors of Ref. \cite{dfa}. Since 1994, hundreds of papers, which
 analyze fractal properties of  time series with the DFA, have been published. In summary DFA works as follows.
Given a time sequence $\{\xi_i\}$ ($i=1,~...,~N$), the DFA is
based upon  the following steps. First,  the entire sequence of
length $N$ is integrated, thereby leading to
\begin{equation}\label{DFAint31}
x(t)=\sum _{i=1}^{t}\left(\xi_i-<\xi> \right),
\end{equation}
where
\begin{equation}\label{DFAint32}
<\xi>=\frac{1}{N}\sum _{i=1}^{N} \xi_i~.
\end{equation}
Second, the time series is divided into $int(N/l)$ non-overlapping
boxes. The number $l$,  which indicates the size of the box, is an
integer smaller than $N$.  A local trend is defined for each box
by fitting  the data in the box.   The linear least-squares fit
may be done with a polynomial function of order $n \geq 0$
\cite{stanley32}. Let $x_l(t)$ be the local trend built with boxes
of size $l$.  Third, a detrended walk is defined as the difference
between the original walk and the local trend given by the linear
least-squares fit according to the following relation
\begin{equation}\label{dfadetr33}
X_{l}(t)=x(t)-x_{l}(t)~.
\end{equation}
Finally, the mean squared displacement of the detrended walk is calculated, that is,
\begin{equation}\label{dfafinavar}
F_{D}^{2}(l)=\frac{1}{N}\sum _{t=1}^{N} [X_{l}(t)]^2~.
\end{equation}
The application of this method of analysis to a number of complex systems (see, for instance,  Refs. \cite{dfa,stanley32} )
shows that
\begin{equation}\label{DFAscarel3}
F_{D}(n)\propto l^{H}.
\end{equation}
Again, according to the traditional wisdom of the methods based on variance,
the exponent $H$ is considered to be a scaling exponent. 
Thus the extent of the departure from the ordinary condition of Brownian motion
is given by $|H- 0.5|>0$.

\subsection{SWA}

The SWA  is a new and powerful method  for
studying the fractal properties of variance \cite{percival}.
SWA decomposes the sample variance of a
time series on a {\it scale-by-scale} basis.  Wavelet Transform makes use of scaling wavelets    that have
the characteristics of being localized in space and in frequencies.
 The wavelets are characterized by the fact that
they can  be localized in the space and depend upon a scaling
coefficient that gives the width of the wavelet. Two typical
wavelets widely used are the Haar wavelet 
and the Mexican hat wavelet \cite{percival}.
A width coefficient $\tau$ defines the scale analyzed by
the wavelet.
 Given a signal $\xi(u)$, the Continuous Wavelet Transform  is defined by
\begin{equation}\label{cwtdhjj3}
W(\tau, t)=\int\limits_{-\infty }^{\infty } \tilde{\psi}_{\tau ,t}(u)~\xi(u)~du~.
\end{equation}
The original signal can be recovered from its Continuous Wavelet Transform via
\begin{equation}\label{invcwtxu9}
\xi(u)=\frac{1}{C_{\tilde{\psi}}} \int\limits_{0}^{\infty} \left[ \int\limits_{-\infty }^{\infty } W(\tau,t) ~\tilde{\psi}_{\tau,t}(u)~dt \right] ~\frac{d\tau}{\tau^2}~.
\end{equation}
Finally, it is possible to prove that
\begin{equation}\label{finrelcwt}
\int\limits_{-\infty }^{\infty } \xi^2(u)~du= \frac{1}{C_{\tilde{\psi}}}\int\limits_{0}^{\infty }\left[\int\limits_{-\infty }^{\infty } W^2(\tau,t)~dt \right]\frac{d\tau}{\tau^2} \equiv \int\limits_{0}^{\infty } S_W(\tau)~d\tau~.
\end{equation}
The function $W^2(\tau,t)/\tau^2$ defines an energy density
function  that decomposes the energy across different scales and
times. Eq. (\ref{finrelcwt}) is the wavelet equivalent to the
Fourier Parseval's theorem. The function  $S_W(\tau)$, defined by
Eq.(\ref{finrelcwt}), is the
wavelet spectral density function that   gives the contribution to
the energy  at the scale $\tau$. 

From Ref. \cite{percival}, we derive that SWA applied 
to studying a noisy seqeuence $\{\xi_i\}$ , at the scale $\tau$,
yields
\begin{equation}\label{scalswdf460}
S_W(\tau)\propto \tau^\alpha .
\end{equation}
The exponent  $\alpha$ is related to the variance scaling exponent $H$ in the same
way as in the conventional
Fourier Analysis. Therefore, $\alpha=2H-1$ for the 
SWA of the noise, and $\alpha=2H$ for the
SWA of the integral of the  noise.
The connection with the methods of scaling detection based on variance is evident.

\section{Artificial sequence analysis}

In this section we verify the theoretical predictions of the previous sections about  the pdf scaling exponent $\delta$ and the variance
scaling exponent, $H$, by using artificial sequences of five million data. With the help of artificial time series, we compare
the methods of analysis based on variance with the DEA. We prove that the DEA always determines the true scaling $\delta$, whereas the  variance based methods detect the true scaling only in the Gaussian case. Thus, in the L\'{e}vy case, only the DEA reveals the true scaling. 
The accuracy of all the numerical points is estimated to be of about $1\%$, for all the cases but the short-time regime of the highly subdiffusional regime studied in Fig.1 and Fig. 2. In this case the error is estimated to be slighltly  larger, of about $2\%$. To make easier for the reader
to interpret the figures, we have not reported the error bars.

\subsection{Gauss statistics: Fractional Brownian diffusion}

Fractional Brownian diffusion is produced by Fractional Brownian
noise.  We generate a time series $\{  \xi_{H,t} \}$ of N data by
using the original algorithm by  Mandelbrot, illustrated in  the book
of Feder \cite{feders}. Our choice is motivated only by historical reasons,  and it is beyond the purpose of this paper to discuss
the computational relevance of other algorithms, more recently proposed to generate FBM \cite{percival,stanley6}. Chosen a value of $H\in [0:1]$, let $\{
\theta_i\}$ be a set of Gaussian random variables with unit
variance and zero mean. The discrete fractional Brownian increment
is given by
\begin{eqnarray*}
\label{fbifeder}
&&\xi_{H,t}= x_H(t)-x_H(t-1) = \\
 &&\frac{n^{-H}}{\Gamma(H+0.5)} \left\{ \sum_{i=1}^{n} i^{H-0.5} ~ \theta_{1+n(M+t)-i}+ \right. \\
 &&\left. \sum_{i=1}^{n(M-1)}\left( (n+i)^{H-0.5}-i^{H-0.5} \right)~ \theta_{1+n(M-1+t)-i}     \right\} , 
\end{eqnarray*}
where $M$ is an integer that should be moderately large, and $n$
indicates the number of the fractional steps for each unit time. Note that the time $t$  is discrete.  In the simulation,   good results are
obtained with $M=1000$ and $n=10$. The time series $\{  \xi_{H,t}
\}$ is then used for generating a diffusion process with the
trajectories (\ref{positions}).

According to the theoretical arguments of Section 3, we expect  $\delta = H$ in this case. To confirm this expectation by means of the statistical analysis we generate five different sequences, with the following values of $H$:  (1) $H=0.8$, (2) $H=0.6$, (3) $H=0.5$, (4) $H=0.4$, (5) $H=0.2$. We analyze these sequences with the SDA (Fig. 1) and with the DEA (Fig.2). The results of the numerical analysis fully confirm our expectation. Let us see all this in more detail. For illustration convenience, in Fig. 1 
we show the normalized standard deviation defined by
\begin{equation}\label{sdvar55}
NSD(l)= \frac{SD(l)}{SD(1)}~.
\end{equation}
where  $SD(l)$ is defined by Eq.(\ref{varvar33}).
With this choice. at $l = 1$ all the numerical results yield, in the ordinate axis, the same value, equal to the unity. 
For the same reason, in Fig. 2, we plot the entropy difference $S(l)-S(1)$, thereby making all five numerical curves depart from the same ordinate value at $l=1$. In both figures the straigh lines are the results of a fitting procedure, based on $f_{SD}(l) = l^{H} + K_{SD}$ in Fig. 1, and on
$f_E(l) = \delta \ln (t) + K_{E}$, in  Fig. 2. 
Note that only the levels of the straight lines of these two figures, namely the values $K_{SD}$'s and $K_{E}$'s, whose actual value is of no interest, are determined by the fitting procedure.  The  parameters $H$ of the straight lines of Fig. 1  are given by the earlier mentioned set of values, selected to create the five artificial sequences. The  parameters $\delta$ of the straight lines of Fig. 2 
are the corresponding scaling values,  prescribed by the theoretical remarks of Section 3, namely  $\delta = H$ for each curve. The
extremely good agreement between theory and numerical experiment,  shown in Fig. 1 and Fig. 2, proves that, as expected, in the case of FBM, the condition $H=\delta$ really holds true.

It is remarkable that for almost all the  values of $H$ the parameters $K_{SD}$ and $K_{E}$ are very close to zero.  This is a consequence of 
a peculiar property of FBM. In general, the statistical analysis of times series is affected by the transition from dynamics  to thermopdynamics. The short-time regime is a kind of dynamic regime and the scaling regime is a kind of thermodynamic regime. It takes time to make a transition from the dynamic to the thermodynamic regime. This property is made especially evident by the analysis of L\'{e}vy walk \cite {giacomo}. In the ideal case of FBM, however, the transition time is expected to be equal to zero. This means  that, in the case of  ideal FBM,  the parameters $K_{SD}$ and $K_{E}$ should vanish. This constraint is satisfactorily fulfilled by all the curves of Fig. 1 and Fig. 2, but those corresponding to $H=\delta = 0.2$. This seems to be a consequence of the numerical inaccuracy, enhanced by the highly antipersistent condition of this case.

\subsection{L\'{e}vy statistics: flight and walk diffusion}

We generate five time series of numbers,
$\{r_i\}$, distributed according to the following inverse power
law
 \begin{equation}\label{l9evywalkdis}
\psi(r)=(\mu-1)\frac{T^{\mu-1}}{(T+r)^\mu}~.
\end{equation}
We select the following values of $\mu$: (1) $\mu=2.8$, (2) $\mu=2.6$, (3) $\mu=2.5$, (4) $\mu=2.4$ and (5) $\mu=2.2$. 
We also select the numbers $s_{i}$, equal to $1$ or to $-1$, randomly according to the coin tossing. We generate the L\'{e}vy flight
of Section 3B, by building up the new sequence $\{\xi_i\}$, where $\xi_{i} \equiv s_{i} r_{i}$.
As to the L\'{e}vy walk, we derive it, as explained in Section 3C, by sewing together into a single sequences many patches. These patches are filled with either an integer number of $1$'s or an integer number of $-1$, according to the coin tossing prescription. The distribution density of the patch length is the same as that of Eq. (\ref{l9evywalkdis}). As a consequence, the L\'{e}vy flight and the L\'{e}vy walk, in the asymptotic time limit have the same scaling, given by
Eq.(\ref{deltalfmkk}).  However, as explained in Section 3C, the L\'{e}vy walk is expected to 
result in a given $H$, predicted by Eq.(\ref{wrongscaling}). 
In Table I we have reported for the reader conveniences the values of $\delta$ and $H$ associated to each $\mu$ of the five artificial sequences that we are considering for our numerical check.

Fig. 3 shows the DEA at work on the time series generating L\'{e}vy flight. The straight lines are fitting functions of the form $f_{E}(l) = \delta \ln (t) + K_{E}$.  The values of the parameters $\delta$, are not fitting parameters, but are  determined by  the theoretical prescription of Table I. 
Figs. 4 and 5 illustrate the results of the SDA and RRA, respectively, applied to the same five time series of Fig. 3. 
Both
figures yield for $H$ a value independent of
$\mu$. This value is $H=0.5$ in all cases. According, to the traditional wisdom, 
this would suggest the wrong conclusion that we are in the presence of
ordinary Brownian motion. We are not, and the DEA is warning us from making this wrong conclusion.
The reason for this misleading result is that these techniques are determined by both the
finite value of the variance, due to statistical limitation and the memoryless nature of the sequence $\{r_i\}$.
The smaller the parameter $\mu$, the smaller the variance, as shown by Fig. 4.
The RRA eliminates this spreading, due to the fact that it normalizes the data by dividing by the standard deviation.

Figs. 6-10 refer to the time series generating L\'{e}vy walk. 
Fig. 6 illustrates the result of the DEA. As in Fig. 3 the straight lines are fitting functions of the form $f_{E}(l) = \delta \ln (t) + K_{E}$, and, again the parameters $\delta$ are those reported in Table I, and are determined by theoretical prescription $\delta =1/(\mu-1)$. 
Figs. 7, 8, 9 and 10 illustrate the results of RRA, 
 SDA, DFA and
 SWA,  respectively. In these cases the straight lines are fitting functions with the form $ f(l) = K l^{H}$. The  parameters $H$  are not fitting parameters, and their values, reported in Table I,  are determined by $H= (4-\mu)/2$.  It is remarkable that, for all these techniques,  the same value of $\mu$ yields the same value of $H$, as the fitting curves show. This fully supports the conclusion that
these seemingly different methods of analysis are actually equivalent, and that all of them are reliable only in the FBM case. On the other hand, 
we notice that the values of $\delta$ and the values of $H$ reported in Table I fit the condition of Eq.(\ref{relHdelta34}) and this a strong evidence that the statistics generated by the time series
is L\'{e}vy statistics. This means that the disagreement between the DEA and the ordinary techniques of analysis can be used for the important purpose of defining the  nature of statistics generated by strange kinetics.

\section{ Significance of the results obtained}

This paper affords a compelling evidence that the DEA is the
only method leading in all conditions to the detection  
of the correct scaling exponent $\delta$. In the case of a sequence of random numbers, that according
to the GCLT should result in an anomalous scaling, the popular Hurst method would lead to the
wrong conclusion that the process observed is equivalent to ordinary 
Brownian motion.
All the traditional methods would lead to quite
correct conclusion only in the case of Gaussian statistics,
a condition which does not mean, of course,
ordinary Brownian diffusion, 
as made evident by the FBM theory of Mandelbrot.
It is also evident that these traditional methods ought
not to be abandoned, even if they have to be used with caution.
The results of Section 6B prove  that the departure
of $\delta$ from $H$ is a clear indication of the occurrence of
L\'{e}vy statistics.  More in general, the 
departure of the traditional methods from the DEA finding might be
used to shed light on statistics as well as on dynamics. Paraphrasing the title of a recent paper\cite{sokolov}, "Do strange kinetics imply unusual thermodynamics?", we can say that one of the basic problems concerning complex systems is that of establishing
if anomalous diffusion (strange kinetics) is compatible or not with ordinary Gaussian distribution (ordinary thermodynamics). The connection between thermodynamics and statistical equlibrium does not need any explanation. Here we are subtly assuming that scaling can be perceived
as a form of thermodynamic equilibrium. If we look to the results of this paper from within this perspective,
we can conclude that FBM is an example of strange kinetic  compatible with ordinary thermodynamics.
We can thus conclude that the joint use of DEA and techniques of analysis based on variance
can assess when strange kinetics force the complex system to depart from ordinar thermodynamics.



\onecolumn

\begin{figure}
\epsfig{file=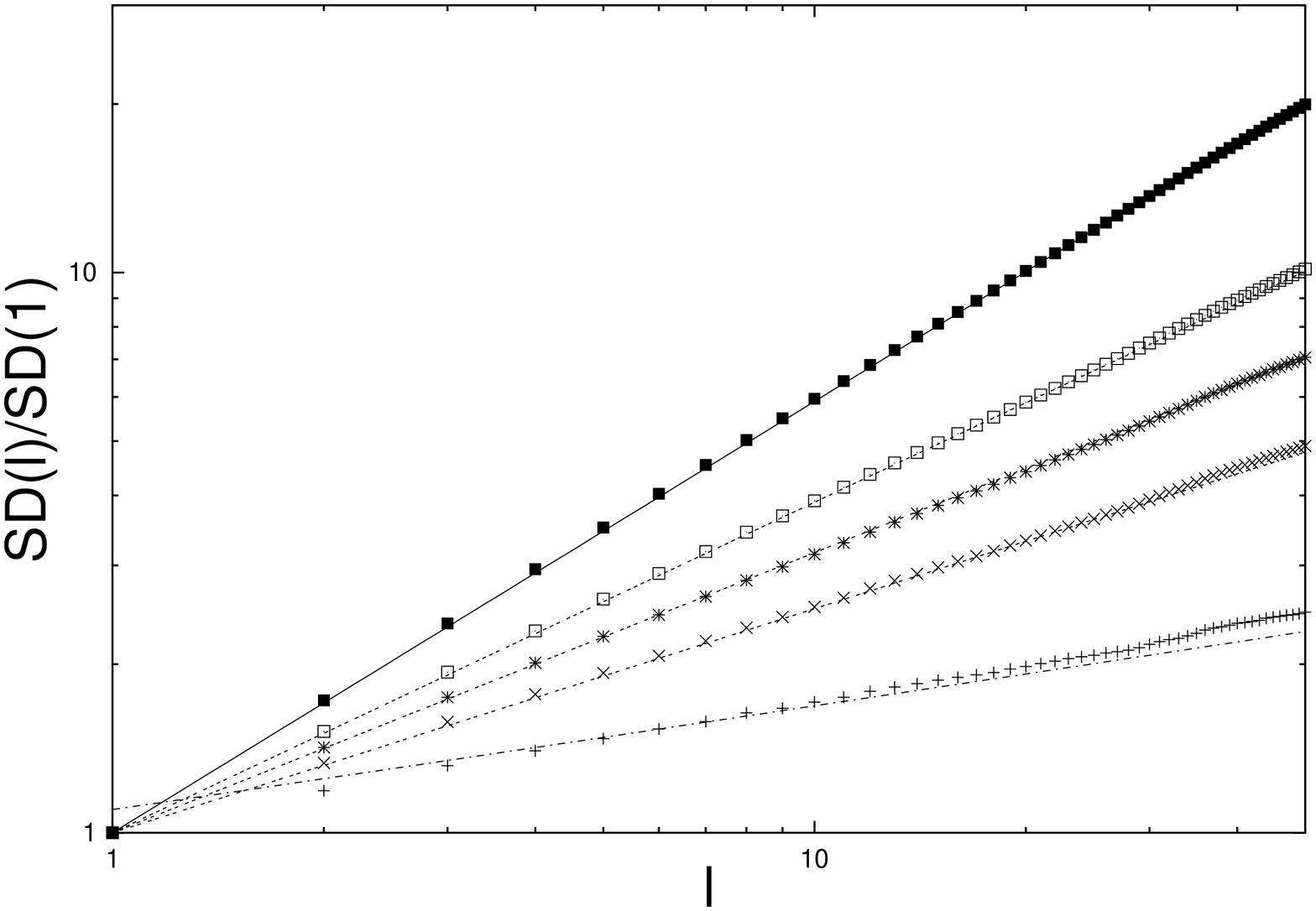,height=16cm,width=12cm,angle=-90}
\caption{SDA in action on the five time series of Fractional Brownian noise of Section 6A. 
In ordinate we plot
 $SD(l)/SD(1)$ as a function of $l$. The straight lines are fitting functions with the form
$f_{SD}(l) = l^{H} + K_{SD}$. 
The values of $H$ are not the result of the fitting procedure, but they correspond to the
set of values used to create the five artificial sequences of Section 6 A.  From the top to the bottom these values are:  (1) $H=0.8$; (2) $H=0.6$; (3) $H=0.5$; (4) $H=0.4$; (5) $H=0.2$. }
\end{figure}

\newpage

\begin{figure}
\epsfig{file=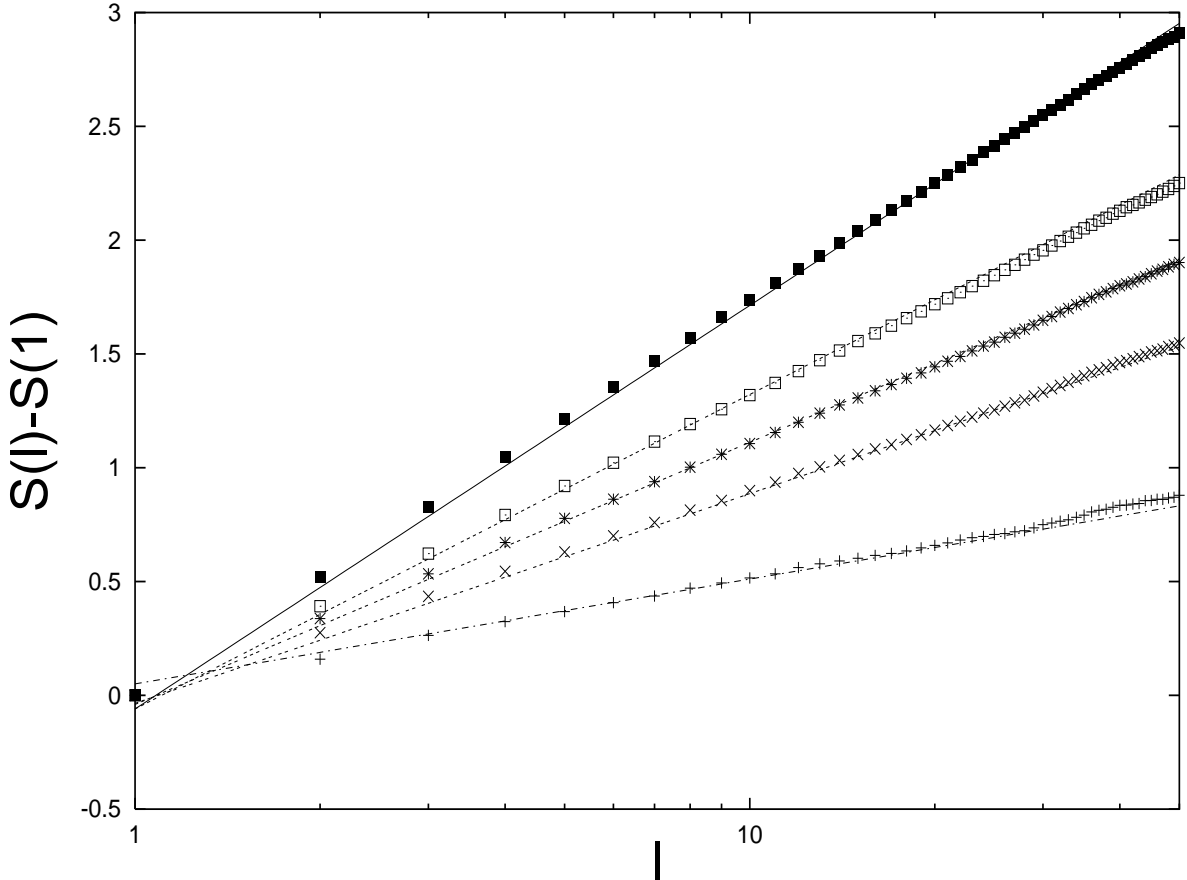,height=16cm,width=12cm,angle=-90}
\caption{DEA of the five time series of Fractional Brownian noise of Section 6A. For illustration convenience, in ordinate we
plot, as a function of $l$, the entropy increment $S(l)-S(1)$.   The
straight lines are fitting functions with the form
$f_{E}(l) = \delta \ln (t) + K_{E}$. The values of $H$ are not the result of the fitting procedure, but they correspond to the
set of values used to create the five artificial sequences of Section 6 A, according to the theoretical prescription $\delta = H$. From the top to the bottom these values are:   (1) $\delta=0.8$; (2)
$\delta=0.6$; (3) $\delta=0.5$; (4) $\delta=0.4$; (5) $\delta=0.2$.}
\end{figure}

\newpage
 \begin{figure} 
\epsfig{file=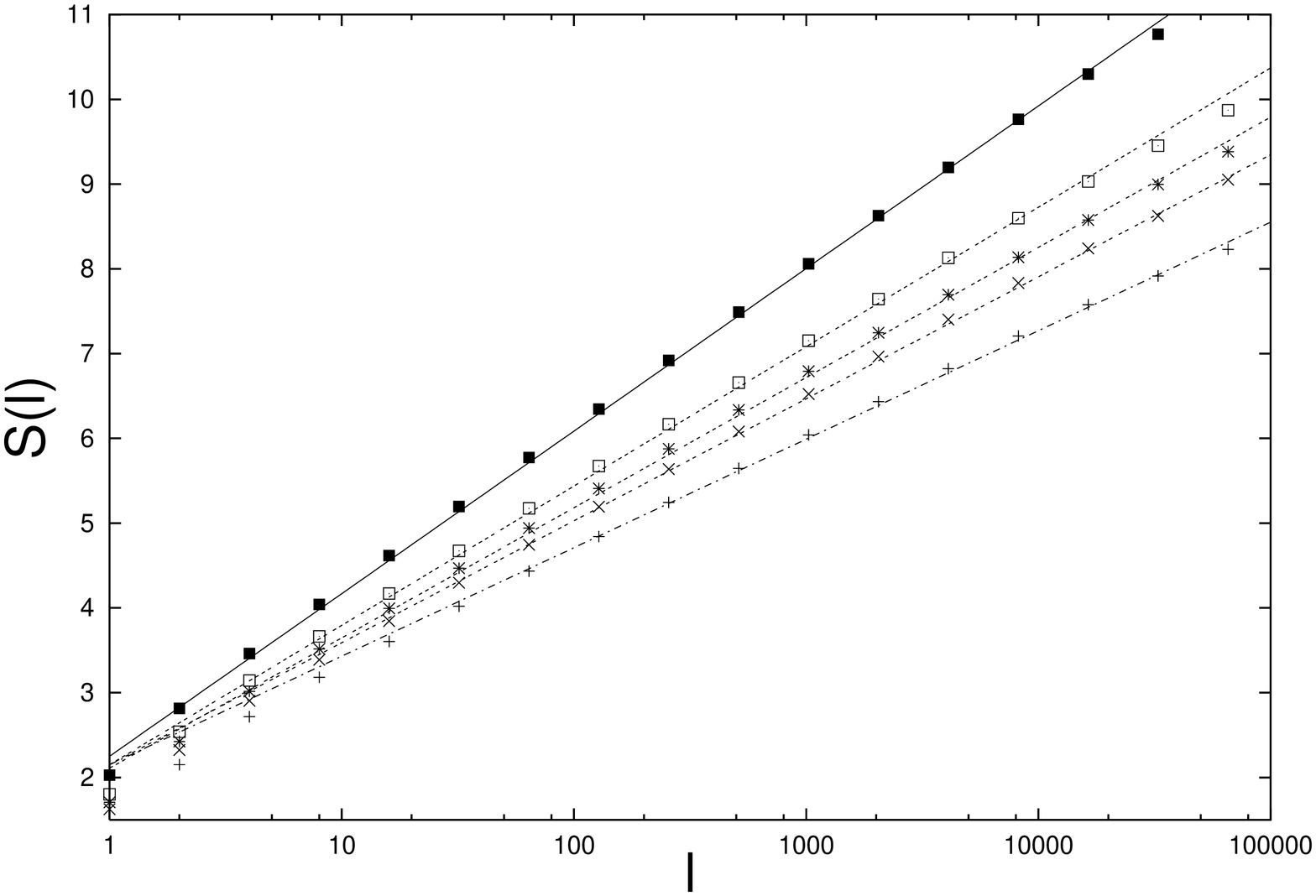,height=16cm,width=12cm,angle=-90} 
\caption{DEA of  the five L\'{e}vy flight time series of Section 6B.
The straight lines are fitting functions with the form $f_{E}(l) = \delta \ln (l) + K_{E}$.
The values of $\delta$ are not the result of the fitting procedure, but they correspond to the
set of values $\{\mu\}$, used to create the five artificial sequences of Section 6 B, according to the theoretical prescription $\delta = 1/(\mu -1)$. 
From the top to the bottom these values, and  the corresponding fitting parameters $K_{E}$, are:
(1) $\delta=0.833, K_{E}= 2.25$; (2)
$\delta=0.714, K_{E} = 2.15$; (3) $\delta=0.667, K_{E} = 2.11$; (4) $\delta=0.625, K_{E} =2.15 $; (5) $\delta=0.556, K_{E} =2.15$.}
\end{figure}

\newpage
 \begin{figure} 
\epsfig{file=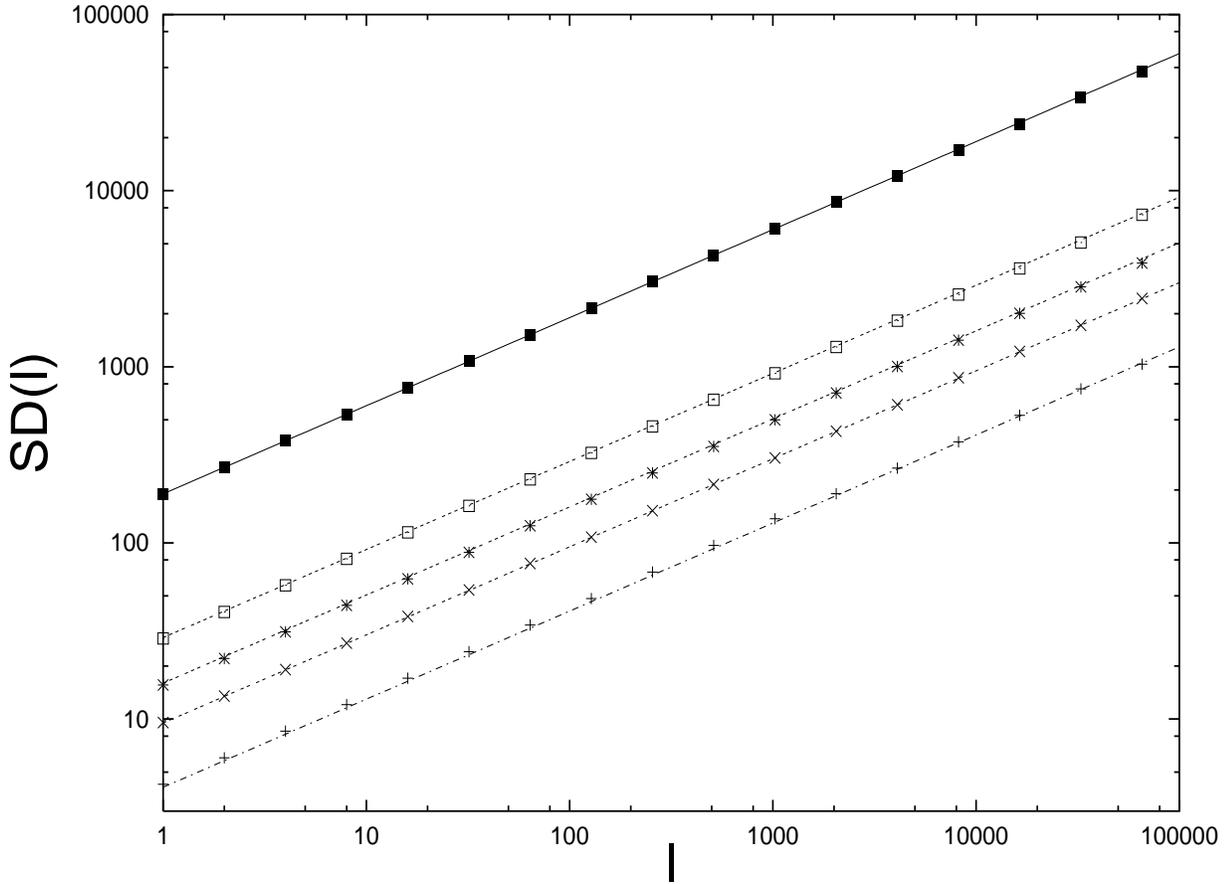,height=16cm,width=12cm,angle=-90}
\caption{ SDA of the five L\'{e}vy flight time series of Section 6B.   
The straigth lines are fitting functions with the form $f_{SD}(l) = K_{D}~ l^{H}$. From the top to the bottom we have: 
(1) $ H =0.5, K_{D}= 190$; (2)
$H =0.5, K_{D} = 29$; (3) $H=0.5, K_{D} =16$; (4) $H= 0.5, K_{D} =9.5 $; (5) $H =0.5, K_{D} =4.1$.}\end{figure}

\newpage
 \begin{figure} 
\epsfig{file=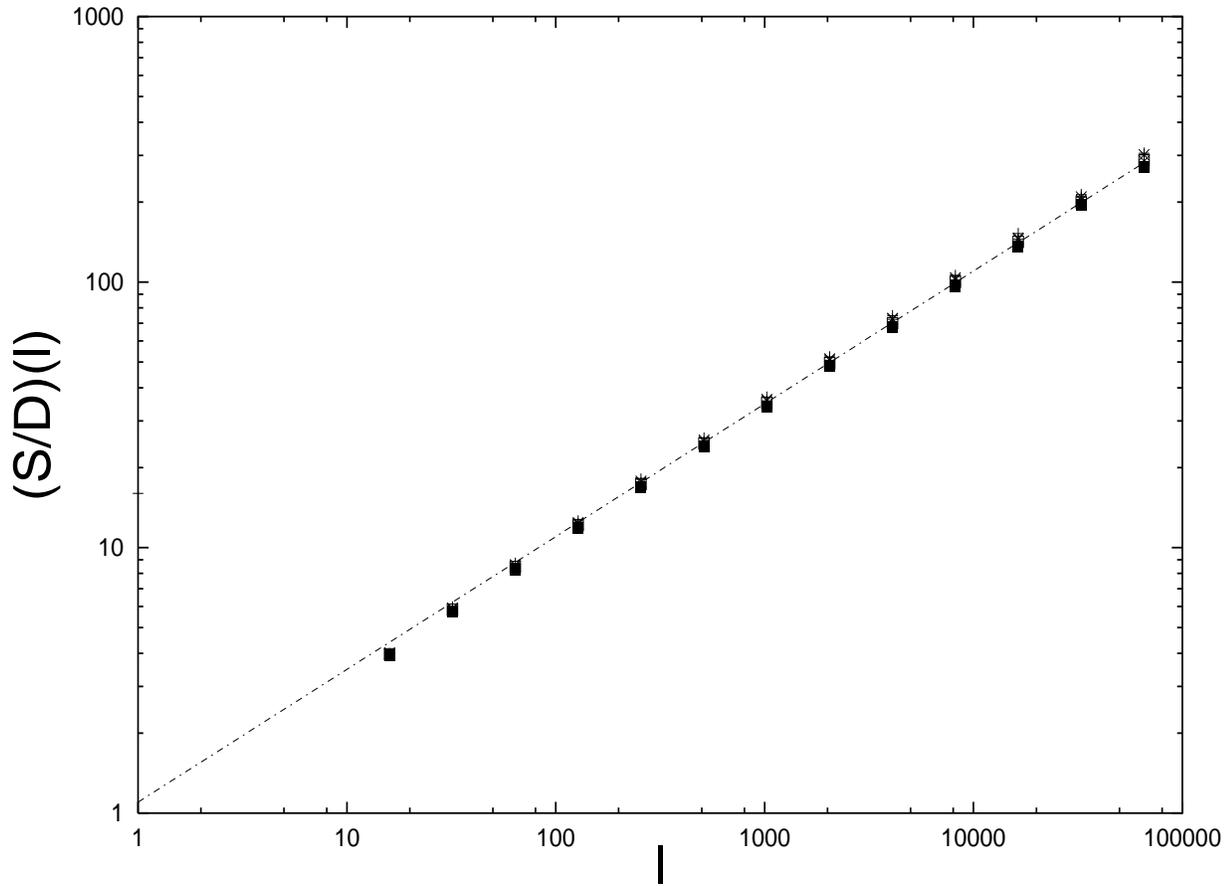,height=16cm,width=12cm,angle=-90}
\caption{ RRA of the five L\'{e}vy flight time series of Section 6B.   All the five cases are fitted by the straight line corresponding to the fitting function $f_{S/D}(l) = K_{S/D} ~l^{H}$, with
$K_{S/D} = 1.1$ and $H = 0.5$. }
\end{figure}

\newpage
 \begin{figure} 
\epsfig{file=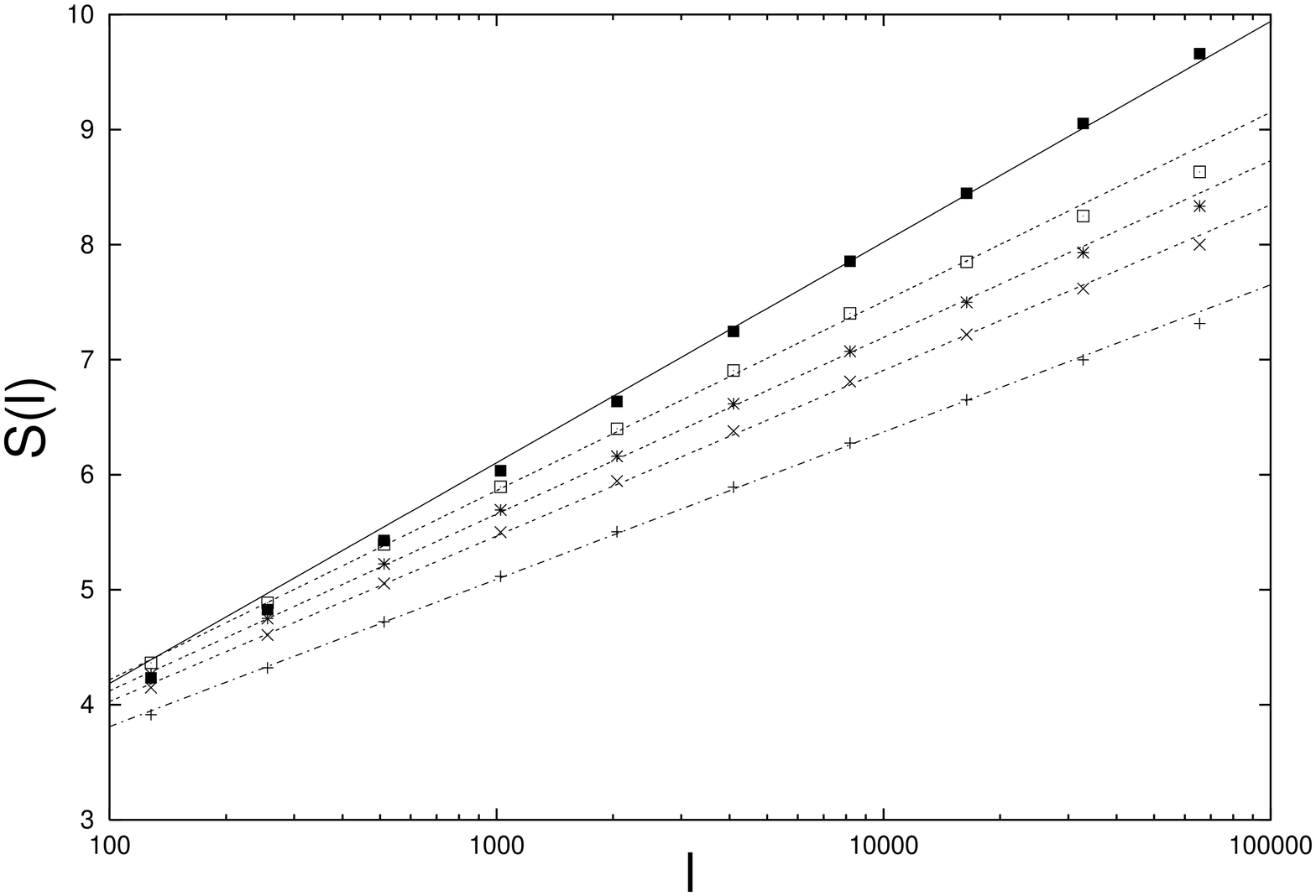,height=16cm,width=12cm,angle=-90} 
\caption{DEA of the five L\'{e}vy walk time series of Section 6B. The straight lines are fitting functions with the form $f_{E}(l) = \delta \ln (l) + K_{E}$.
The values of $\delta$ are not the result of the fitting procedure, but they correspond to the
set of values $\{\mu\}$, used to create the five artificial sequences of Section 6B, according to the theoretical prescription $\delta = 1/(\mu -1)$. 
From the top to the bottom these values, and  the corresponding fitting parameters $K_{E}$, are: (1) $\delta=0.833, K_{E}= 0.35$; (2)
$\delta=0.714, K_{E} = 0.93$; (3) $\delta=0.667, K_{E} = 1.05$; (4) $\delta=0.625, K_{E} =1.15 $; (5) $\delta=0.556, K_{E} =1.25$.}
\end{figure}

\newpage
 \begin{figure} 
\epsfig{file=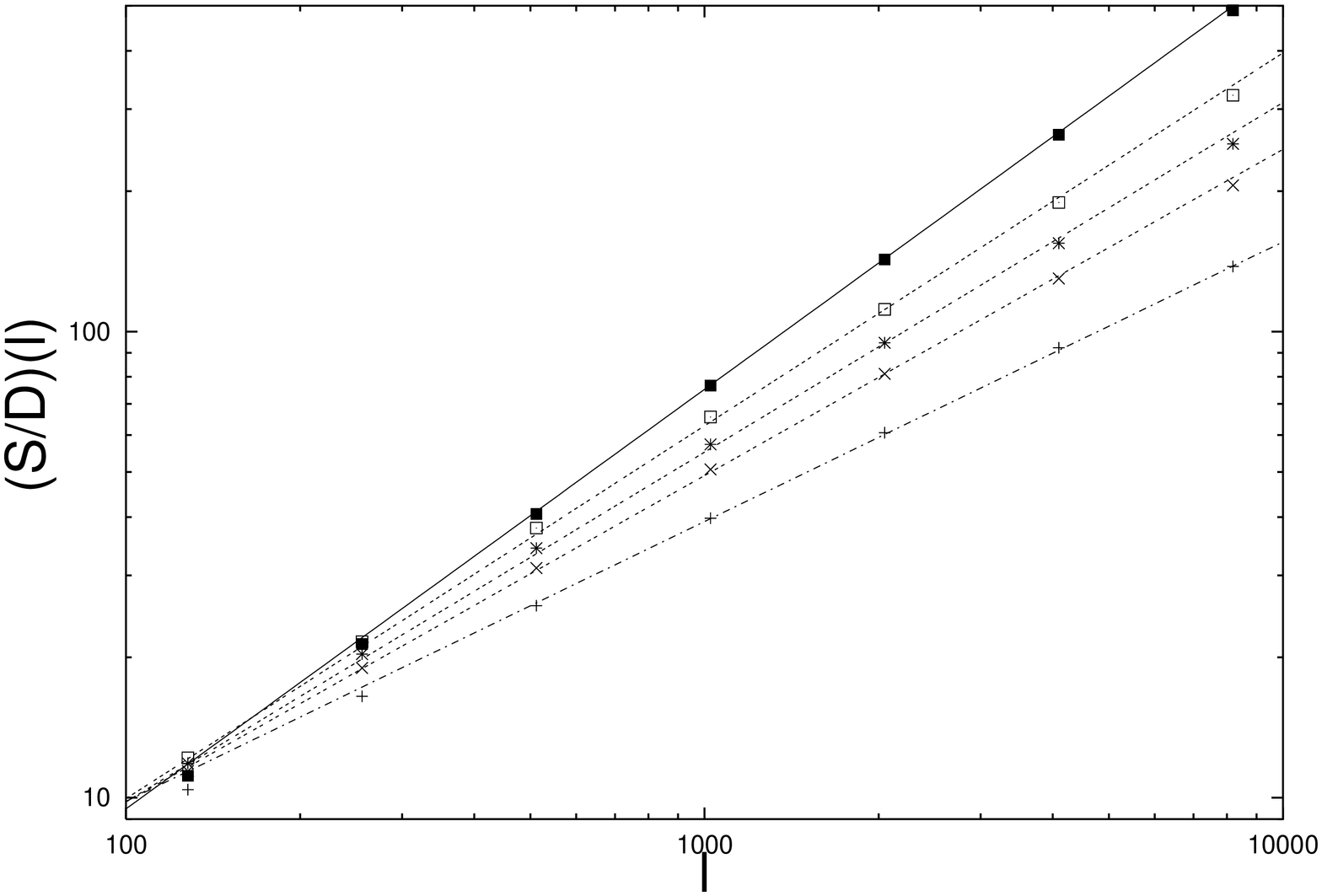,height=16cm,width=12cm,angle=-90}
\caption{ RRA of the five L\'{e}vy walk time series of Section 6B. The straight  lines
are fitting functions with the form  $f_{S/D}(l) = K_{S/D}~ l^{H}$. The values of $H$ are not the result of the fitting procedure, but they correspond to the
set of values $\{\mu\}$, used to create the five artificial sequences of Section 6B, according to the theoretical prescription $H = (4-\mu)/2$. 
From the top to the bottom these values, and  the corresponding fitting parameters $K_{SD}$, are: (1) $H =0.9, K_{S/D}= 0.15$; (2)
$H=0.8, K_{S/D} = 0.25$; (3) $H=0.75, K_{S/D} = 0.75$; (4) $H=0.7, K_{S/D} =0.39 $; (5) $H =0.6, K_{S/D} =0.62$.}
\end{figure}

\newpage
 \begin{figure} 
\epsfig{file=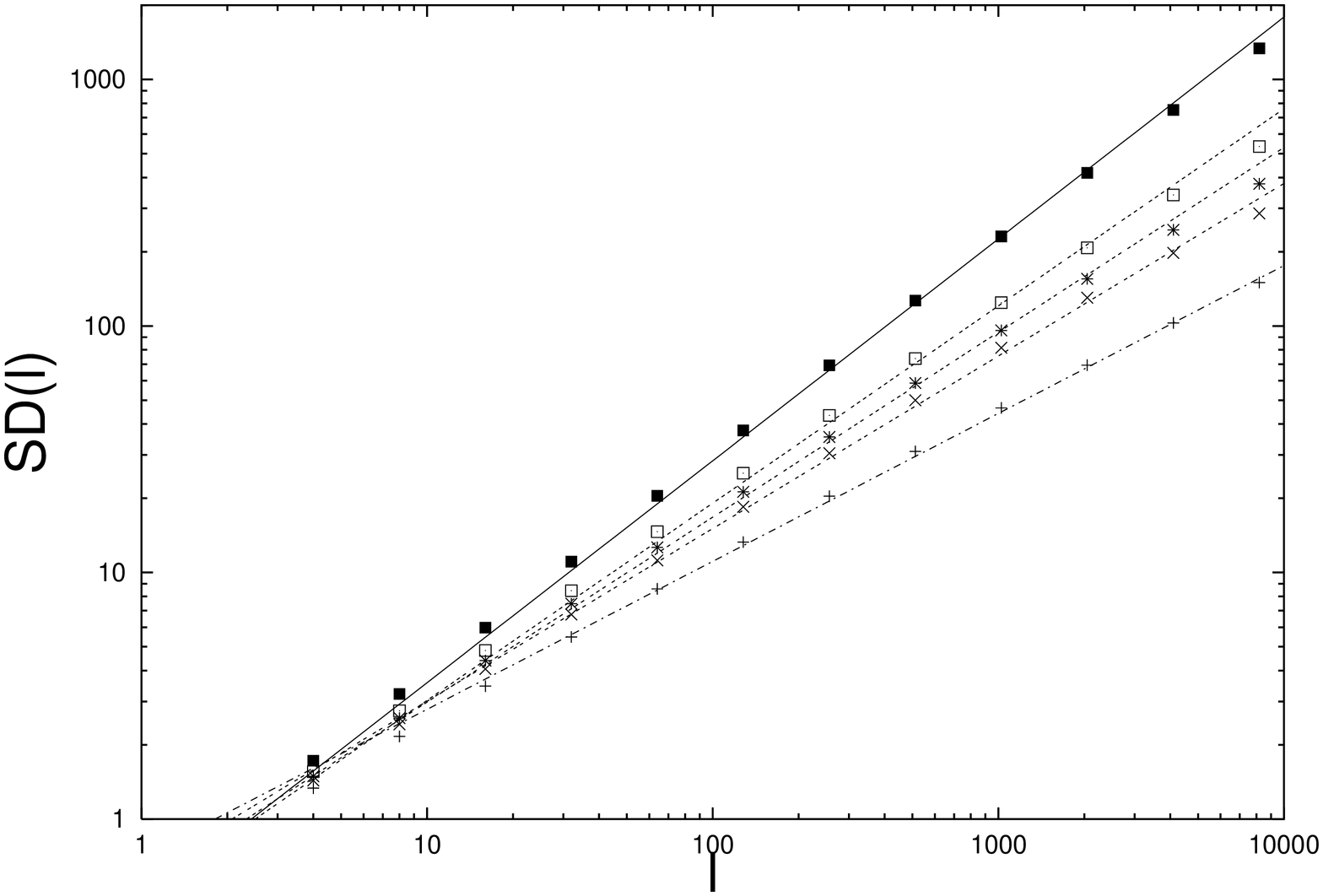,height=16cm,width=12cm,angle=-90}
\caption{ SDA of   the five L\'{e}vy walk time series of Section 6B. 
The straight  lines
are fitting functions with the form  $f_{SD}(l) = K_{SD}~ l^{H}$.The values of $H$ are not the result of the fitting procedure, but they correspond to the
set of values $\{\mu\}$, used to create the five artificial sequences of Section 6B, according to the theoretical prescription $H = (4-\mu)/2$. 
From the top to the bottom these values, and  the corresponding fitting parameters $K_{D}$, are(1) $H =0.9, K_{D}= 0.45$; (2)
$H=0.8, K_{D} = 0.48$; (3) $H=0.75, K_{D} = 0.53$; (4) $H=0.7, K_{D} =0.6 $; (5) $H =0.6, K_{D} =0.7$.  }
\end{figure}

\newpage
 \begin{figure} 
\epsfig{file=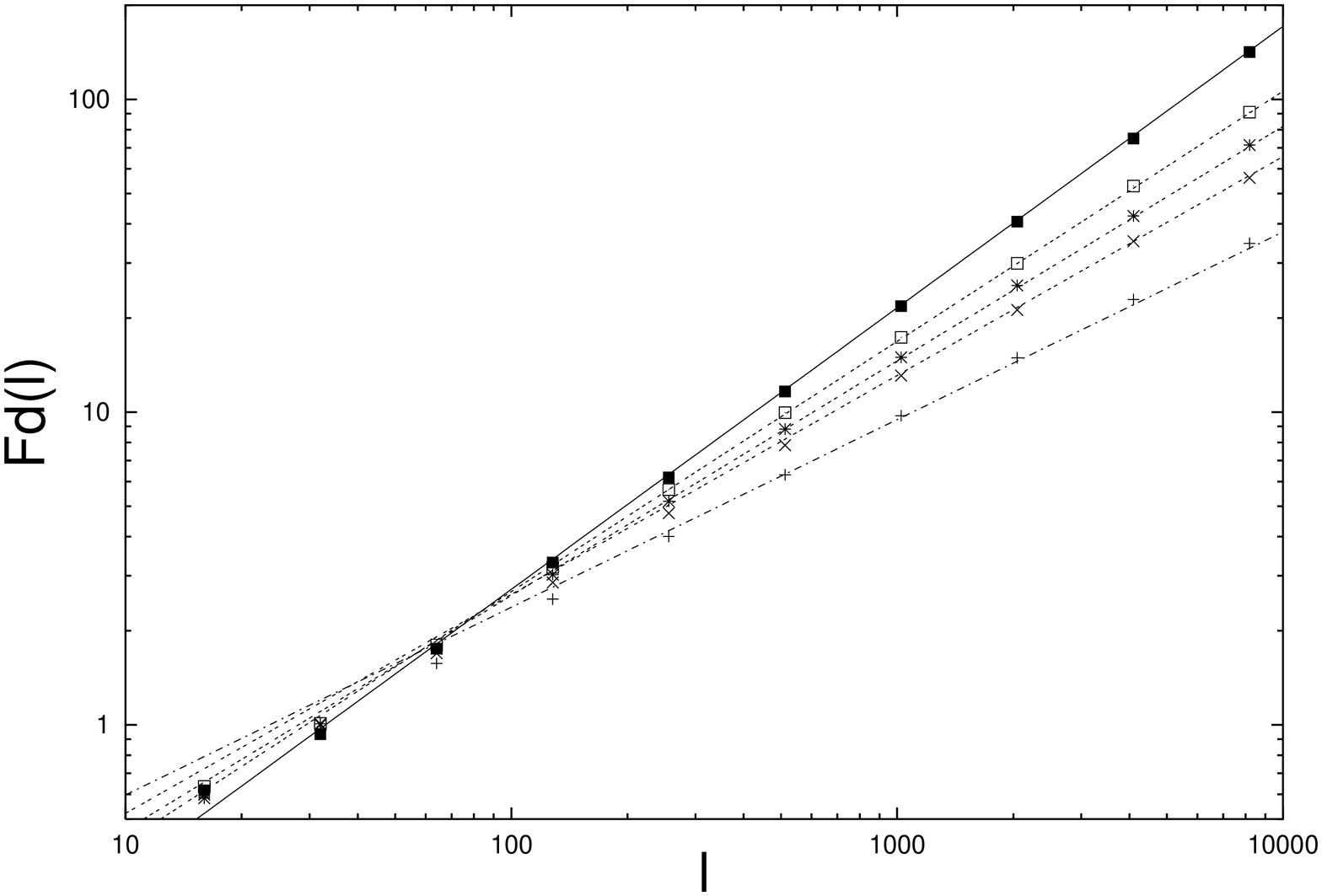,height=16cm,width=12cm,angle=-90}
\caption{ DFA of   the five L\'{e}vy walk time series of Section 6B. 
The straight  lines
are fitting functions with the form  $f_{F}(l) = K_{F}~ l^{H}$. The values of $H$ are not the result of the fitting procedure, but they correspond to the
set of values $\{\mu\}$, used to create the five artificial sequences of Section 6B, according to the theoretical prescription $H = (4-\mu)/2$. 
From the top to the bottom these values, and  the corresponding fitting parameters $K_{F}$, are:  (1) $H =0.9, K_{F}= 0.043$; (2)
$H=0.8, K_{F} = 0.067$; (3) $H=0.75, K_{F} = 0.0.082$; (4) $H=0.7, K_{F} =0.0.104 $; (5) $H =0.6, K_{F} =0.15$.  }
\end{figure}

\newpage
 \begin{figure}
\epsfig{file=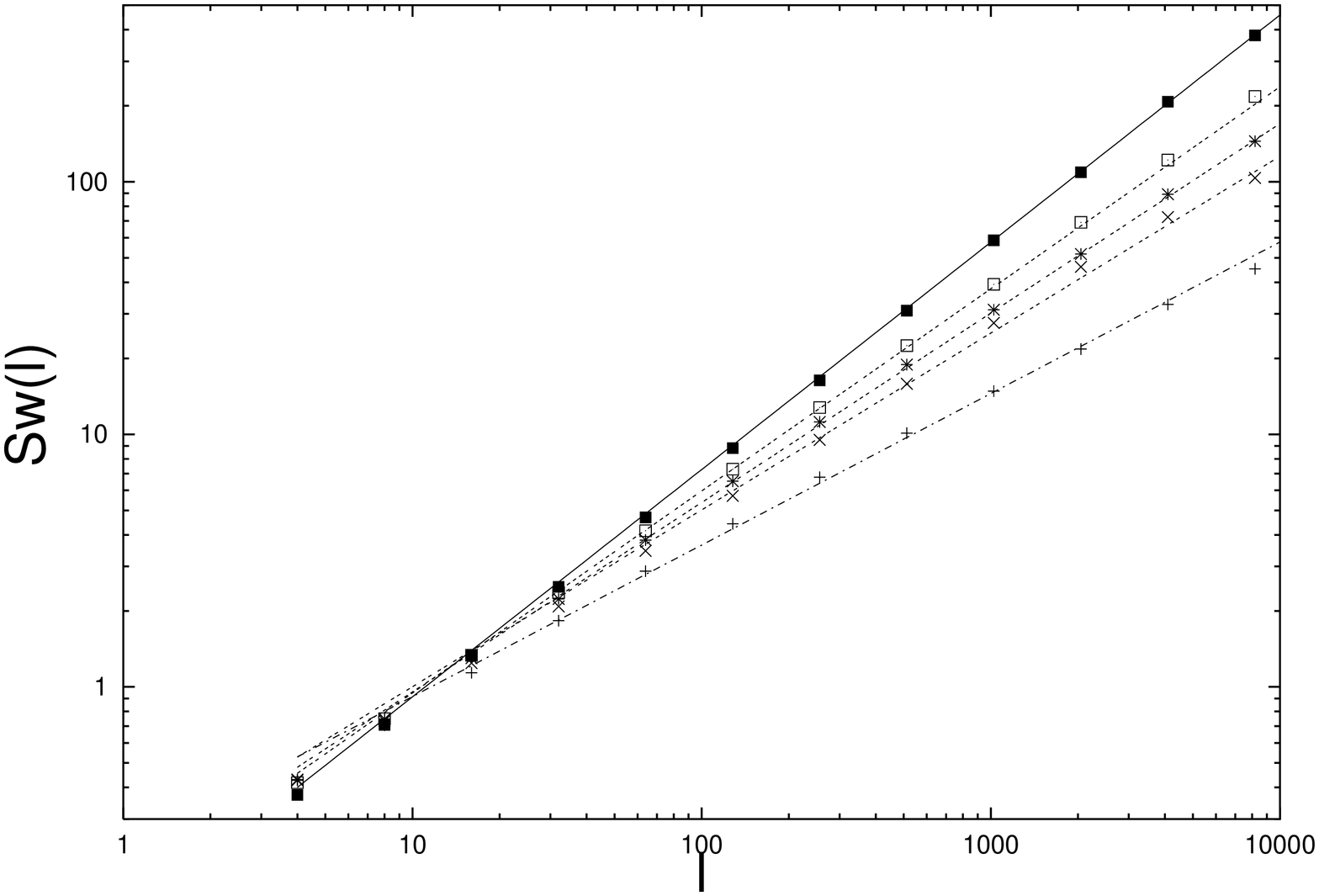,height=16cm,width=12cm,angle=-90}
\caption{ SWA of   the five L\'{e}vy walk time series of Section 6B. 
The straight  lines
are fitting functions with the form  $f_{W}~(l) = K_{W} ~l^{H}$. The values of $H$ are not the result of the fitting procedure, but they correspond to the
set of values $\{\mu\}$, used to create the five artificial sequences of Section 6B, according to the theoretical prescription $H = (4-\mu)/2$. 
From the top to the bottom these values, and  the corresponding fitting parameters $K_{W}$, are: (1) $H =0.9, K_{W}= 0.115$; (2)
$H=0.8, K_{W} = 015$; (3) $H=0.75, K_{W} = 0.17$; (4) $H=0.7, K_{W} =0.2 $; (5) $H =0.6, K_{W} =0.23$.
The wavelet spectral density is calculated using the Maximun Overlap Discrete Wavelet Transform with the Daubechies H4 discrete wavelet [8].}
\end{figure}


\newpage

\begin{table}
\begin{center}
  \begin{tabular}{|c|c|c|}
$\mu$	&	$H$	&	 $\delta$	\\ \hline
2.2	&	0.90	&	0.833	\\ \hline
2.4	&	0.80	&	0.714	\\ \hline
2.5	&	0.75	&	0.667	\\ \hline
2.6	&	0.70	&	0.625	\\ \hline
2.8	&	0.60	&	0.556	
  \end{tabular}
\end{center}
\caption{Theoretical relation between the waiting time
distribution power  exponent $\mu$ and the variance scaling
exponent $H$ and the pdf scaling exponent $\delta$.}
\end{table}

\end{document}